# A self-field electromagnetic behavior of the $MgB_2$ coil analyzed with the A-V formulation-based finite element method

Alexander Chervyakov
*Laboratory of Information Technologies, Joint Institute for Nuclear Research,*
*Joliot Curie 6, 141980 Dubna, Russia*
*acher@jinr.ru*

Abstract

The paper discusses the A-V formulation of Maxwell's equations for modeling superconductors with the finite-element method in three dimensions. The formulation is then used to analyze the electromagnetic behavior of a circular $MgB_2$ coil due to the ramp of the externally applied direct current in self-field conditions. In our modeling setup, the transport current is allowed to directly ramp through the coil volume at a controlled rate. The two gauge conditions are imposed on both potentials and used for their numerical computation, the Coulomb gauge and the temporal gauge. The continuity of the total current density is ensured regardless whether or not the temporal gauge is used to define the electric scalar potential. The nonlinear resistivity of the $MgB_2$ material is represented as a function of the electrical field and is accounted for by fitting parameters and the field-dependence of the critical current density from existing experimental data. We find the current and field distributions during the current ramp time and in the regime of DC stabilization. It is shown that the $MgB_2$ coil is capable of generating the fields below 4 T despite the large values of the applied transport current. The DC voltage-current curve is computed numerically for two different gauge conditions resulting in slightly different values for the coil critical current despite the gauge-independence of the magnetic and electric fields.

## Introduction

Nowadays, superconducting coils are widely used over their conventional counterparts in magnetic systems of various electromagnetic devices due to the higher current capacity of superconducting materials that can be achieved at smaller conductor sizes. Most notable applications include particle accelerators and colliders, nuclear magnetic resonance (NMR) facilities, fusion reactors, wind energy generators and magnetic energy storage systems, rotating electrical machines, magnetic resonance imaging (MRI) magnets and compact particle accelerators for radio, proton and ion therapy [1-3]. Knowing the electromagnetic behavior of superconducting coils is therefore necessary to predict the field and current distributions and optimize the design of these systems. For their numerical modeling, it is however necessary to overcome strong nonlinearities of ferromagnetic and superconducting materials in order to attain sufficient accuracy of calculation for the required uniformity of magnetic fields [4]. Although various numerical techniques for solving Maxwell's equations are currently available for this purpose (see, e.g., [4,5] and references therein), the electromagnetic modeling of superconductors with modern software still needs user-defined customization.

The paper analyzes the electromagnetic behavior of a circular $MgB_2$ coil due to the ramp of a transport direct current in self-field conditions by using the finite-element method for the numerical solution of Maxwell's equations in terms of magnetic vector and electric scalar potentials (A-V formulation) in three dimensions. The A-V formulation is well-known as a fundamental tool for simplifying Maxwell's equations [6-8] and their numerical solutions with the finite-element method [9-12] capable of providing the efficient modeling of various electromagnetic devices, including also superconductors, once the nonlinear properties of superconducting materials are properly accounted for [13-27]. The choice of formulation is appropriate for several reasons. First, the use of the magnetic vector potential is useful to tackle the nonlinearities of magnetic materials surrounding coil. Second, the formulation allows us to access DC characteristics of superconductors, such as the voltage-current (V-I) curve and the effective critical current, as far as a smooth relaxation of the externally applied current followed by its ramp is necessary to determine these quantities [28-31]. Third, the transport direct current (DC) can directly be ramped through the coil volume at a controlled rate by adding its density to the r.h.s. of the equation describing Ampere' law. The total current of a superconductor is then a superposition of the applied current and the internally induced current of the opposite polarity. In accordance with Ampere's law, the divergence-free condition for the total current density is ensured regardless whether or not the temporal gauge is used to define the electric scalar potential.

Despite the transition temperature for the $MgB_2$ material is lower than that for high temperature superconductors (HTS), the material itself is cheaper, lighter weight, possesses better mechanical stability and more homogeneous distribution of the critical current density [32,33]. Furthermore, the $MgB_2$ conductors can be fabricated in form of the round wires thus promising to have the low AC losses when operated below 20 K [34]. However, these conductors have relatively low irreversibility fields and their critical current densities drop rapidly as the magnetic fields increase. For this reason, they are primarily suited for low to medium-field applications typically below 5 T at operating temperatures around 10-20 K, such as compact heavy particle accelerators for medical treatment, electrical motors and generators, MRI magnets, and superconducting magnetic energy storage systems, filling a niche where the cost and cooling simplicity are matters [35, 36].

In this paper, the electromagnetic behavior of a circular $MgB_2$ coil in three dimensions is analyzed with the finite-element software COMSOL Multiphysics [37]. For stability of the numerical computation, the two different gauge conditions are imposed on the potentials and used for their numerical computation, the Coulomb gauge and the temporal gauge. The nonlinear resistivity of the $MgB_2$ material is represented as a function of the electric field and is accounted for by fitting parameters and the field-dependence of the critical current density from existing experimental data [38, 39]. We find the current and field distributions during the current ramp time and in the regime of DC stabilization. It is shown that the $MgB_2$ coil is capable of generating the fields below 4 T despite the large values of the applied transport current. The DC voltage-current curve is computed numerically for two different gauge conditions resulting in slightly different values for the coil critical current despite the gauge-independence of the magnetic and electric fields.

1. A-V formulation for finite-element modeling of superconductors

The numerical solution of Maxwell's equations with the finite-element method in terms of magnetic vector potential $\boldsymbol{A}$ and electric scalar potential $V$ is a well-established approach for the efficient electromagnetic modeling of various electrical devices [9-12] including superconductors, provided that the non-linear properties of superconducting materials are properly accounted for [13-27]. For the latter application, the A-V formulation (also referred to as the A-V-J formulation [17,18] if the main state variables $\boldsymbol{A}$ and $V$ are accomplished by the current density $\boldsymbol{J}$ in order to speed up the convergence of a numerical solution) can be briefly described by using a computational domain $\Omega$ composed of the two complementary regions, the current-carrying region of a superconductor $\Omega_{sc}$ and the remaining non-conducting region $\Omega_{nc} = \Omega/\Omega_{sc}$, for an illustration.

The A-V formulation [6-8] operates with the magnetic vector potential $\boldsymbol{A}$ in the entire problem domain $\Omega$ and the electric scalar potential $V$ in the conducting region $\Omega_{sc}$ to solve the corresponding subset of Maxwell's equations in the quasistatic approximation. The magnetic flux density $\boldsymbol{B}$ and the electric field $\boldsymbol{E}$ are then expressed in terms of these potentials as follows

$$\boldsymbol{B} = \nabla \times \boldsymbol{A}, \tag{1}$$

$$\boldsymbol{E} = -\nabla \cdot V - \partial_t \boldsymbol{A}, \tag{2}$$

where the electric field $\boldsymbol{E}$ is induced up to the gradient of a scalar potential $V$ by a time-varying magnetic vector potential $\boldsymbol{A}$ according to Faraday's law. The bold and regular letters refer to three-component vector and scalar functions, respectively. In particular, the vector and its norm are denoted by the bold and regular variant of the same letter. For each material region encountered in $\Omega$, the magnetic flux density $\boldsymbol{B}$ is further related to the magnetic field strength $\boldsymbol{H}$ via the constitutive relation $\boldsymbol{B} = \mu\boldsymbol{H}$ with $\mu$ being the magnetic permeability of a material. Depending on a material type, its magnitude varies from a constant for the superconducting and air areas to the nonlinear function of $H$ for highly permeable magnetic regions incorporated into $\Omega$. Using this relation together with definition (1) and neglecting displacement currents, Ampere's law can be expressed in terms of magnetic vector potential $\boldsymbol{A}$ by a single double-curl equation

$$\nabla \times \left(\frac{1}{\mu} \nabla \times \boldsymbol{A}\right) = \boldsymbol{J} + \nabla \cdot \psi \ \text{ in } \Omega, \tag{3}$$

where $\psi$ is a Lagrange multiplier introduced to constrain the potential $\boldsymbol{A}$. The current density $\boldsymbol{J}$ reads

$$\boldsymbol{J} = \boldsymbol{E}/\rho(E) + \boldsymbol{J}_e \ \text{ in } \Omega_{sc}, \tag{4}$$

where the first term $\boldsymbol{J}_i = \boldsymbol{J} - \boldsymbol{J}_e$ is the density driven in a superconductor by the electric field (2) and the second $\boldsymbol{J}_e$ is the externally applied current density. The function $\rho(E)$ is the electrical resistivity of a superconducting material

$$\rho(E) = \frac{E_c}{J_c(B)} \left(\frac{E}{E_c}\right)^{1-1/n} \tag{5a}$$

specified by the appropriate values of the field-dependent critical current density $J_c(B)$, the critical electric field $E_c$ and the power exponent $n$. By these equations, the induced electric field (2) can be treated as a dissipative electric field of the form $\boldsymbol{E} = \rho(J_i)\boldsymbol{J}_i$ with the resistivity (5a) being re-expressed in terms of the density $\boldsymbol{J}_i$ as follows

$$\rho(J_i) = \frac{E_c}{J_c(B)}\left(\frac{J_i}{J_c(B)}\right)^{n-1}. \tag{5b}$$

Using equations (2) and (4), the conservation of the total current $\nabla \cdot \boldsymbol{J} = 0$ inherent in Ampere's law (3) is expressed by a double-gradient equation

$$\nabla \cdot (-\nabla V/\rho(E) - \partial_t \boldsymbol{A}/\rho(E) + \boldsymbol{J}_e) = 0 \text{ in } \Omega_{sc} \tag{6}$$

to be solved for the electric scalar potential $V$ together with the ground boundary condition $\mathrm{V} = 0$ imposed on the boundary $\partial\Omega_{sc}$ of a superconductor. In this way, the continuity of the total current is preserved. In addition, we use the Lagrange multiplier in equation (3) to impose the Coulomb gauge condition on the potential $\boldsymbol{A}$ in the entire problem domain to stabilize its numerical computation

$$\nabla \cdot \boldsymbol{A} = 0, \text{in } \Omega. \tag{7}$$

As such, equations (3) and (6) form a system of coupled partial differential equations for the magnetic vector potential $\boldsymbol{A}$ and the electric scalar potential $V$, respectively. To accelerate their computation, it is also advantageous to separate the compution of the total current density (4) from equation (3) by creating the axillary dependent variable $\boldsymbol{J}$. Using this variable, we assume also that the boundary $\partial\Omega_{sc}$ of a superconductor is electrically insulated $\boldsymbol{n} \cdot \boldsymbol{J} = 0$.

Yet another way to exploit gauge freedom existing in the definition of the potentials (1) and (2) is to impose the temporal gauge on the scalar potential $V$ in the conducting region and the Coulomb gauge on the vector potential $\boldsymbol{A}$ in the non-conducting region

$$V = 0, \text{in } \Omega_{sc}, \qquad \nabla \cdot \boldsymbol{A} = 0, \text{in } \Omega_{nc}. \tag{8}$$

By the gauge (8), the potential $V$ becomes totally eliminated from equation (6), whose purpose is now to constrain the numerical solution of equation (3) for the magnetic vector potential $\boldsymbol{A}$ rather than define the electric scalar potential $V$. It is worth noting that the A-V formulation in the temporal gauge (8) still differs from the so-called A-J formulation (see, e.g., [26, 27]) by equation (6) which preserves, even for $V = 0$, the continuity of the total current. Both numerical schemes of the A-V formulation using the gauge conditions (7) and (8) with and without electric scalar potential $V$ respectively, are explored in the next section to analyze the electromagnetic properties of a circular superconducting coil designed as a part of the magnetic system for a particle accelerator.

## 2. A self-field electromagnetic behavior of a circular $MgB_2$ coil

In this section, we analyze the electromagnetic response of a superconducting $MgB_2$ coil to the ramp of a transport DC current in self-field conditions. A three-dimensional model consists of the circular coil placed symmetrically above and below the median plane of the large spherical air domain used for magnetic insulation. The magnetic components surrounding coil are not considered at this stage of computation. The geometry of the model is shown in Fig.1.

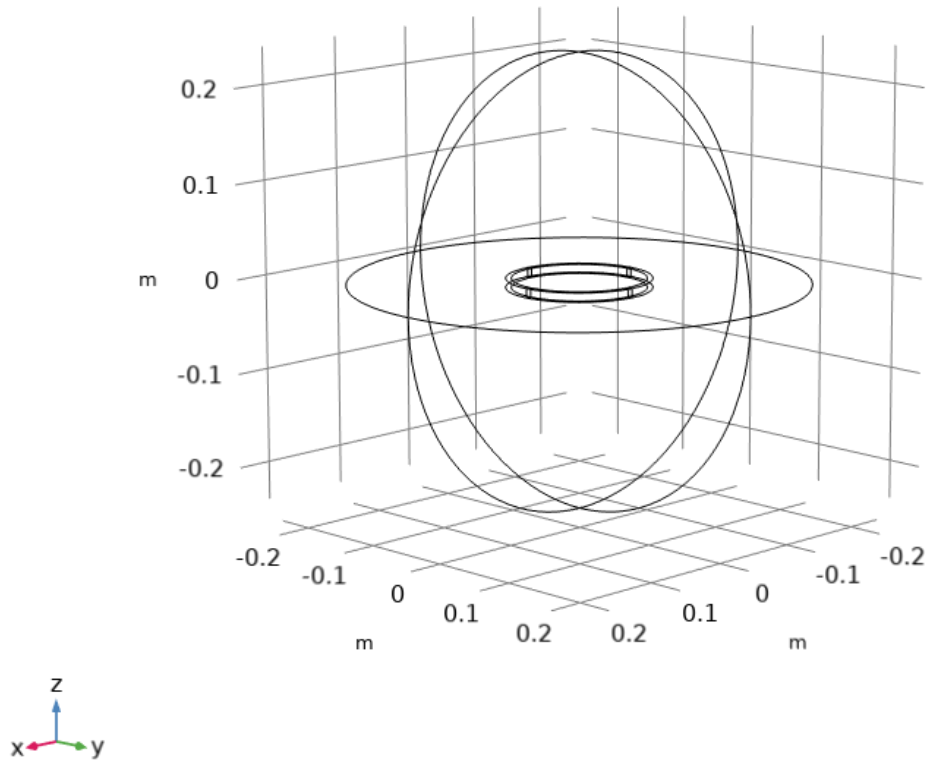

*Figure 1:The geometry of a superconducting coil model used for computation.*

The geometry of a circular coil represents the axisymmetric hollow cylinder formed by rotation around the $z$-axis of the rectangle $(9.60\ \mathrm{mm} \times 5.82\ \mathrm{mm})$ placed over the $(z, x)$-plane with its center at the point $(0, 73.11\ \mathrm{mm})$. The multiturn coil is modeled as a homogenized current carrying body with multiple wires arranged and placed in a potting material. The geometry of the insulating air domain represents a large sphere given the radius of $240\ \mathrm{mm}$.

Following [38], the field dependence of the critical current density $J_c(B)$ for the $MgB_2$ conductor at 20 K is approximated by the exponential law

$$J_c(B) = J_{c0} \exp\left[-(B/B_0)^{3/2}\right], \tag{9}$$

where $B_0 = 0.95\ \mathrm{T}$ and $J_{c0} = 1.71 \cdot 10^9\ \mathrm{A/m^2}$ are the fitting parameters [39]. For this conductor, the values of the critical electric field and the power exponent in equations (5a) and (5b) are specified, respectively as $E_c = 10^{-6}\ \mathrm{V/m}$ and $\mathrm{n} = 100$ [38, 39].

The external current density $\boldsymbol{J}_e$ in equation (4) is taken in the following form

$$\boldsymbol{J}_e = J_e\left\{-y/\sqrt{x^2+y^2},\ x/\sqrt{x^2+y^2}, 0\right\} \tag{10}$$

to satisfy the divergence-free condition $\boldsymbol{\nabla} \cdot \boldsymbol{J}_e = 0$. A smooth ramp function is applied over the time of $1.02 \cdot 10^5$ seconds to inject the density (10) into the coil volume by slowly ramping up its magnitude $J_e$ from zero to the final value of $1.368 \cdot 10^9\ \mathrm{A/m^2}$ representing 80% of the zero-field critical current density $J_{c0}$, while exceeding the curve $J_c(B)$ at fields of $0.4\ \mathrm{T}$ and higher. The corresponding peak value of the applied current amounts to $I_e = 76433\ \mathrm{A}$. Since the produced magnetic field is estimated to be above $2\ \mathrm{T}$, using these values allows superconducting coil to reach its critical current. After reaching its peak value during the ramp time, the density (10) continues to propagate under DC conditions until full dissipation of the Faraday electric field (2) at the time of $5.5 \cdot 10^5$ seconds. The field-induced current flows in the opposite direction to the ramped current, trying to decrease effectively the growth of the total current by creating its own magnetic field according to Lenz's law (see Fig.2). The value of the induced current can be regulated by adjusting the rate of the ramped current. Increasing the present rate of $0.76\ \mathrm{A/s}$, for example, by two orders of magnitude to the value of $76.43\ \mathrm{A/s}$ would make the induced current approximately equal to the applied current, with the zero total current at the ramp start. Since the induced current disappears after the full dissipation of the Faraday electric field (2) under the relaxed DC conditions, choosing this rate would take much longer stabilization time and therefore does not be considered here.

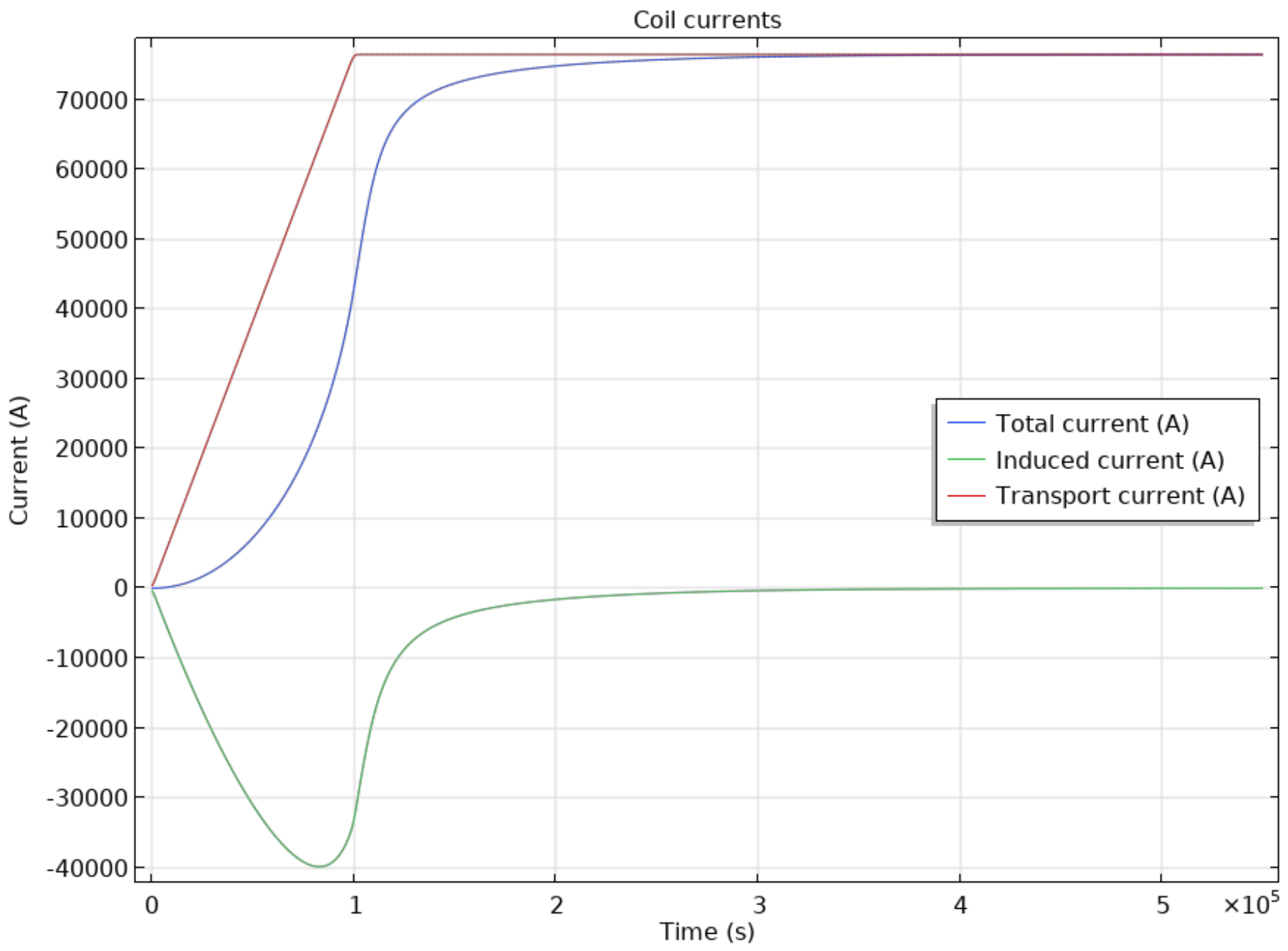

*Figure 2: The applied transport current reaches its peak value 76433 A in 1.02E5 seconds. At the time of 5.5E5 seconds the total current achieves the value of 76425 A as compared to the value 76433 A of the applied current meaning that the induced current is reduced to a small value of -8.19 A. The speed of the ramp is* $0.76\ A/s$.

Under stabilized DC conditions, where the Faraday-induced electric field (2) completely vanishes, the total coil current reaches the peak value of the applied current to flow persistently with no electric field in a superconductor (see Fig.2). The DC current, which drives the terminal voltage across the coil length to a value above the average electric field criterium of $10^{-6}\ \mathrm{V/m}$, is chosen to define the effective critical current of the coil.

The electromagnetic behavior of the $MgB_2$ coil is analyzed using the finite-element software COMSOL Multiphysics [37]. The quantities of interest are the distributions of the magnetic field, the current density and the electric scalar potential as well as the DC characteristics of the $MgB_2$ coil such as the current-voltage curve and the effective critical current. The fully coupled solution of the A-V formulation is derived for magnetic vector potential $\boldsymbol{A}$ by using the Magnetic Fields interface, whereas for electric scalar potential $V$ and the total current density $\boldsymbol{J}$ by using the appropriated PDE modules.

The calculations with and without the electric scalar potential are carried out by applying the gauge conditions (7) and (8), respectively, while the conservation of the total current is preserved by equation (6). For these calculations, we use the same conforming mesh with 30662 finite elements generated and optimized to ensure the convergence and stability of numerical solutions at reasonable quality. The scalar function $V$ is approximated by the Lagrange elements up to fifth order, while the vectorial functions $\boldsymbol{A}$ and $\boldsymbol{J}$ by the edge finite elements up to third order. The direct PARDISO solver based on the multi-frontal factorization of stiffness matrices is used to achieve the fast convergence of both solutions. The gauge fixing for $\boldsymbol{A}$-field node is activated alongside with this solver to impose either the gauge (7) for the entire problem domain, or the gauge (8) for superconducting and air regions, thus illuminating possible instabilities and/or singularities of the algebraic systems. The absolute and relative tolerances are set at $10^{-3}$ and $10^{-6}$, respectively.

The computation using gauge (7) consumes 11.2 GB of the physical memory and 4754 seconds of the computation time, while using the gauge (8) - 9.4 GB of the physical memory and 3439 seconds of the computation time. The both result in similar distributions of the magnetic field and the current density inside the $MgB_2$ coil shown in Figs. (3)-(6).

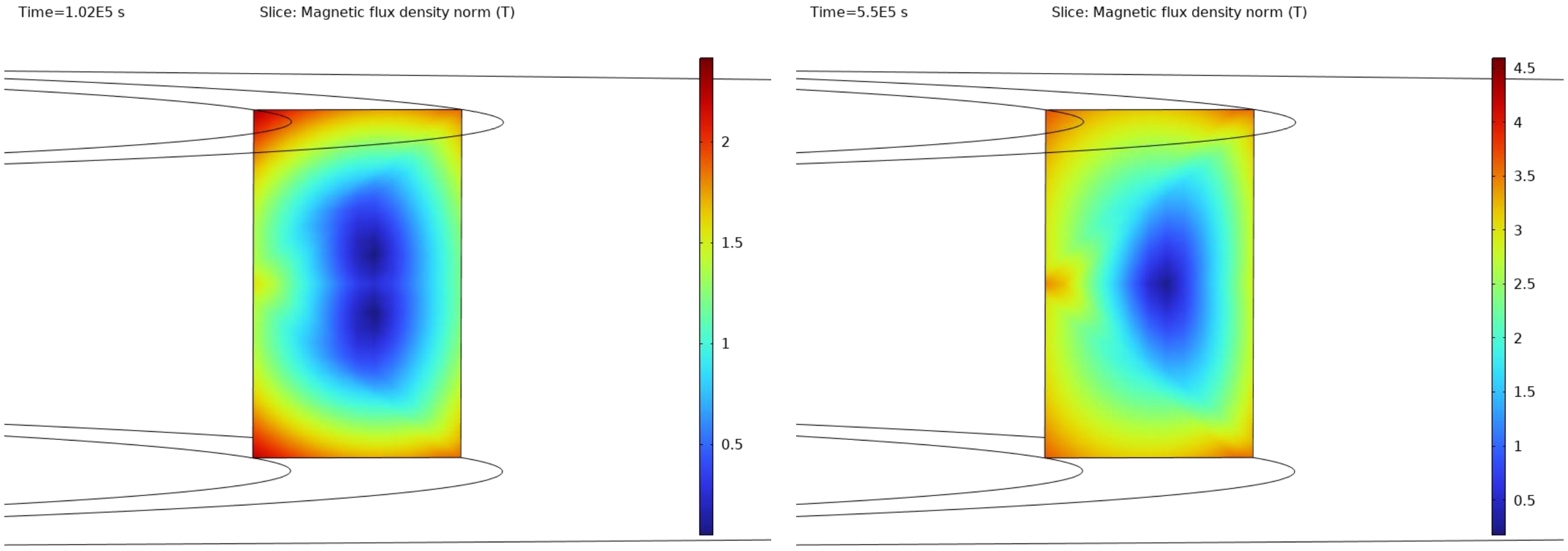

*Figure 3: Magnetic field distributions over the coil cross-section at the end of the ramp time (left) and the DC stabilization (right).*

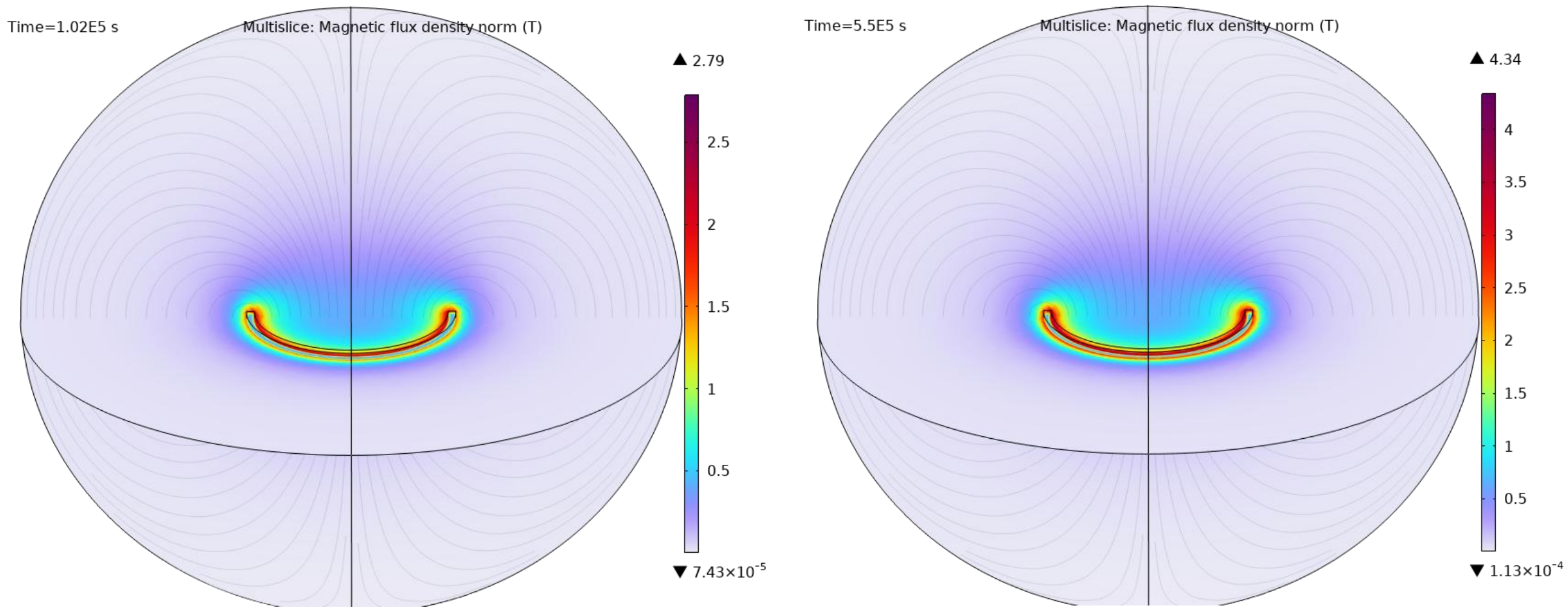

*Figure 4: Magnetic field distributions inside and around the coil volume at the end of the ramp time (left) and the DC stabilization (right).*

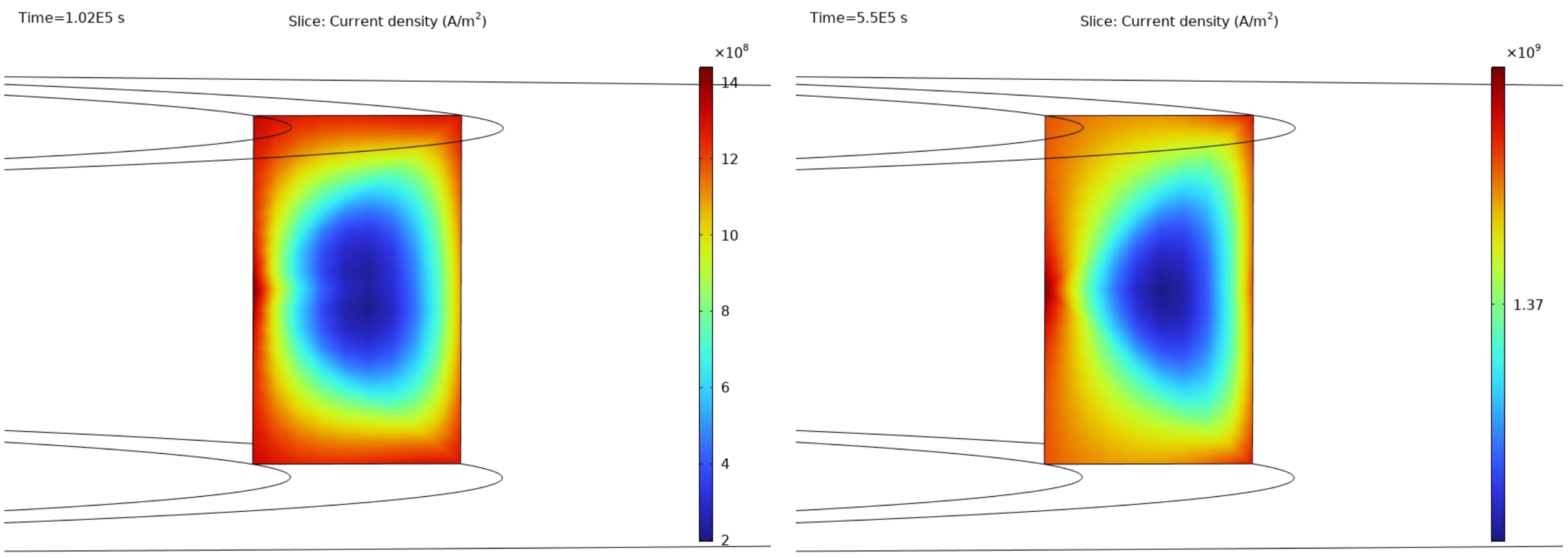

*Figure 5:Current density distributions over the coil cross-section at the end of the ramp time (left) and the DC stabilization (right).*

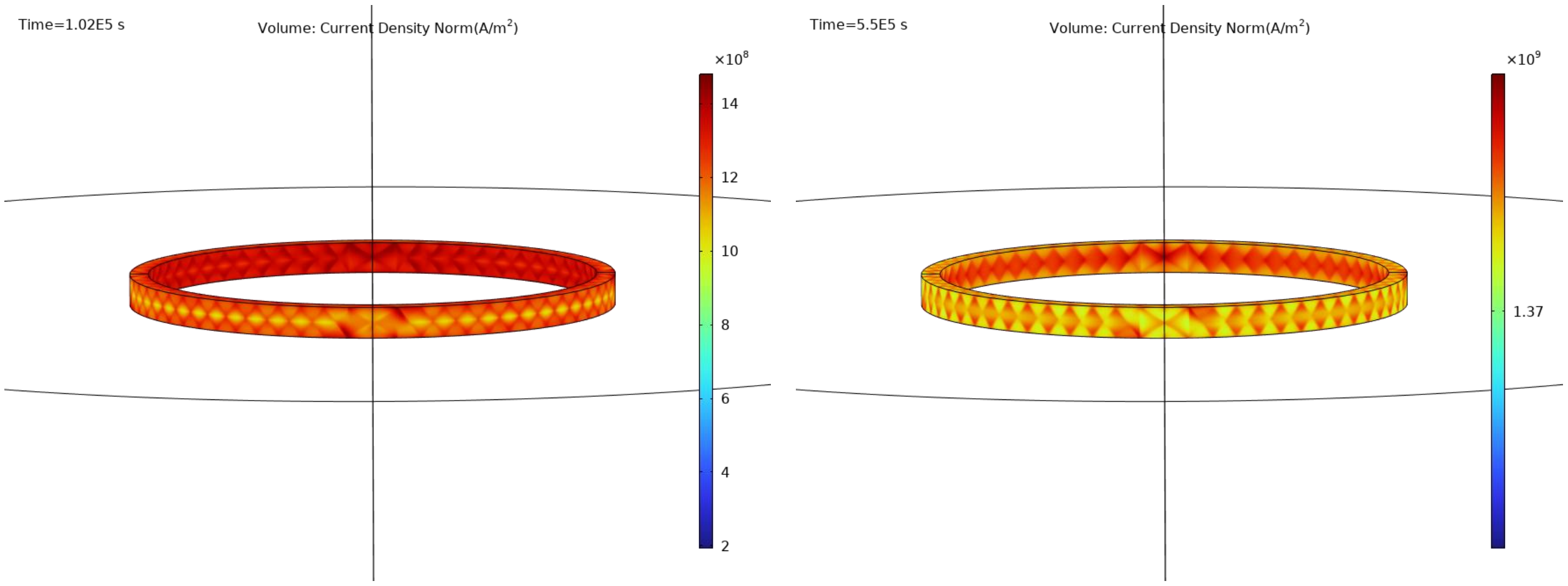

*Figure 6: Current density distributions inside the coil volume at the end of the ramp time (left) and the DC stabilization (right).*

The current and the field distributions shown in Figs. (3)-(6) exhibit the typical behavior for superconductors (see, e.g., presentation [19]), but the $MgB_2$ coil is only capable to generate the moderate magnetic fields below 4 T even for high values of applied current.

Additionally, the computation with the gauge (7) provides us with a numerical solution of the equation (6) for the electric scalar potential grounded on the coil surface. This yields a small contribution to the Faraday's electric field shown in Fig.7.

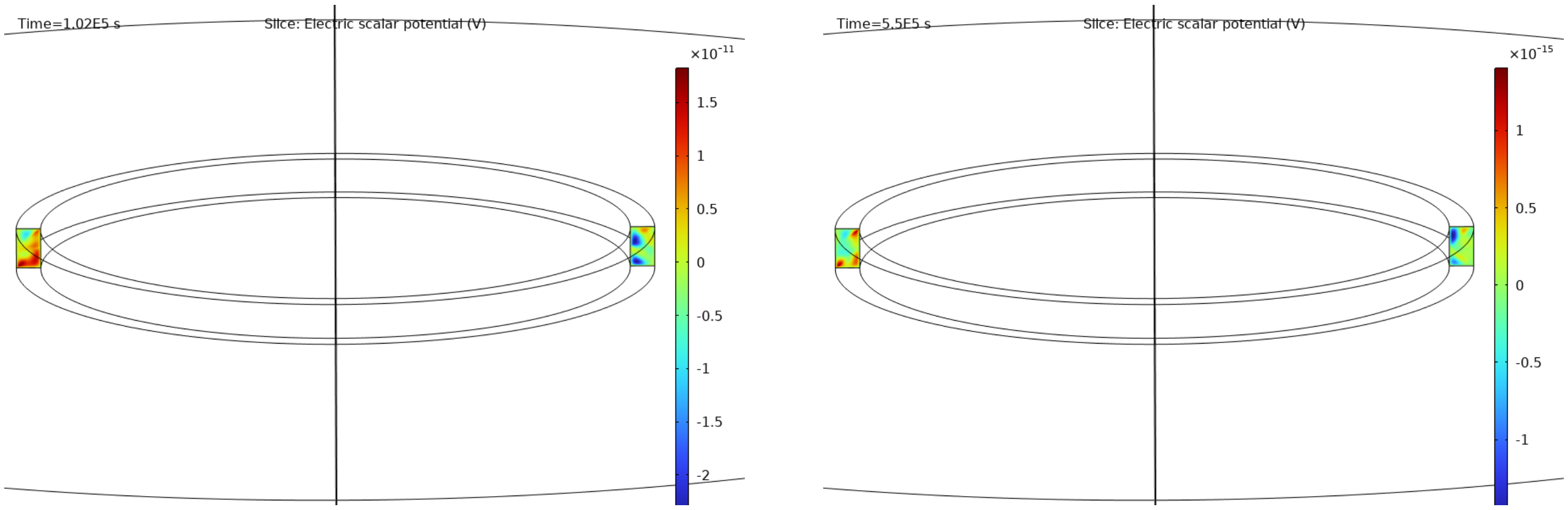

*Figure 7: Distribution of the electric scalar potential over the coil cross-section at the end of the ramp time (left) and the DC stabilization (right).*

The present modeling setup, in which the monotonically ramped transport current drives the terminal voltage across the coil length to a value above the average electric field criterium of $10^{-6}$ V/m, is used to define the effective critical current of the coil. As described above, we ramp up the applied transport current $I_e$ from zero to the peak value of 76433 A in $1.02 \cdot 10^5$ seconds and then keep it a constant until the time of $5.5 \cdot 10^5$ seconds to compute the electric field averaged over the coil cross-sectional area along the entire coil length as a function of the total current at each time step. The results of calculations with gauge conditions (7) and (8) are shown in Fig.8.

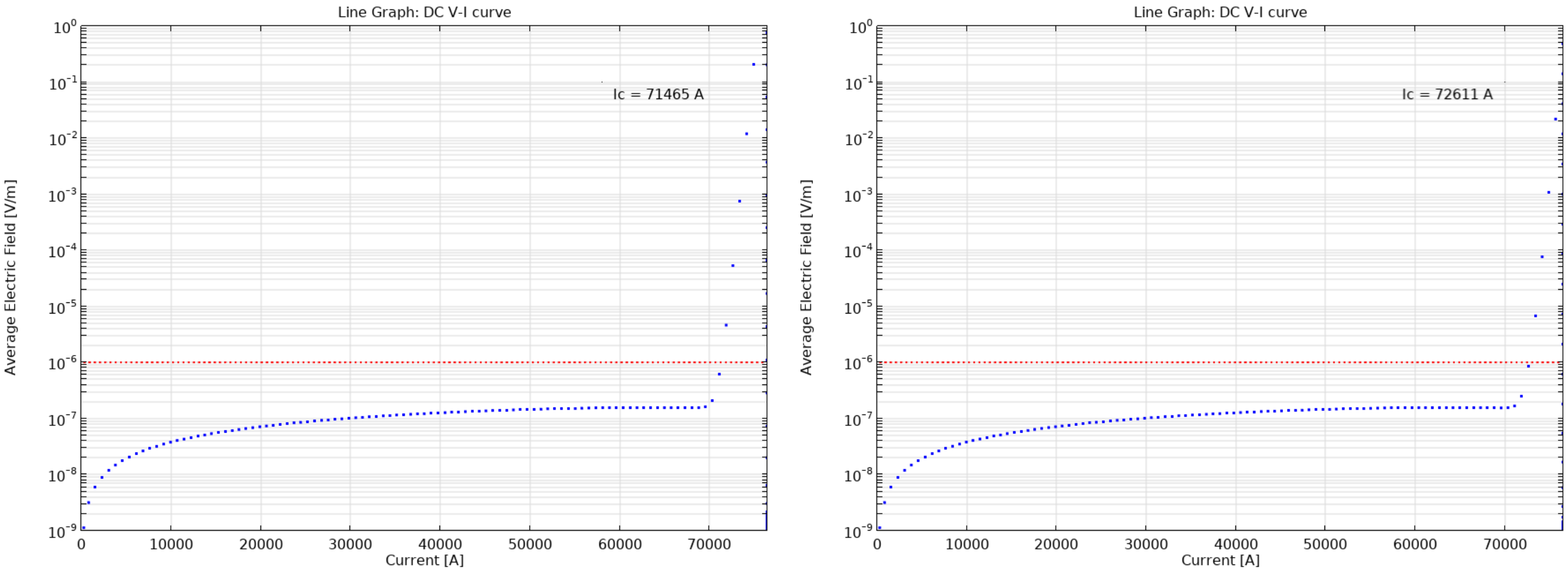

*Figure 8: The averaged over the coil cross-section electric field calculated along the entire coil length as a function of the applied current obtained in calculation using the gauge conditions (7) (left) and (8) (right).*

The results of both numerical computations with conditions (7) and (8) lead to the slightly different values for the coil critical current despite the gauge-independence of the magnetic and electric fields in the analytical sense.

## Conclusion

The paper discusses the A-V formulation of Maxwell's equations for modeling superconductors with the finite-element method in three dimensions. The formulation is then used to analyze the electromagnetic behavior of a circular $MgB_2$ coil due to the ramp of the externally applied direct current in self-field conditions. The modeling of superconductors in terms of magnetic vector and electric scalar potentials is useful to tackle the nonlinearities of magnetic materials surrounding the coil and to access DC characteristics of superconductors, such as the voltage-current (V-I) curve and the effective critical current. Furthermore, the transport direct current can directly be ramped through the coil volume at a controlled rate by adding its density to the r.h.s. of the equation describing Ampere' law. The total current of a superconductor is then a superposition of the applied current and the internally induced current of the opposite polarity. In accordance with Ampere's law, the divergence-free condition for the total current density is ensured regardless whether or not the temporal gauge is used to define the electric scalar potential.

The electromagnetic behavior of a circular $MgB_2$ coil in three dimensions is analyzed with the finite-element software COMSOL Multiphysics. For stability of the numerical computation, the two different gauge conditions are imposed on the potentials and used for their numerical computation, the Coulomb gauge and the temporal gauge. The nonlinear resistivity of the $MgB_2$ material is represented as a function of the electric field and is accounted for by fitting parameters and the field-dependence of the critical current density from existing experimental data. We find the current and field distributions during the current ramp time and in the regime of DC stabilization. It is shown that the $MgB_2$ coil is capable of generating the fields below 4 T despite the large values of the applied transport current. The DC voltage-current curve is computed numerically for two different gauge conditions resulting in slightly different values for the coil critical current despite the gauge-independence of the magnetic and electric fields.

## Acknowledgements

The computational support from HybriLIT Heterogeneous Computing Platform (MLIT, JINR) is acknowledged.

## References

1. L. García-Tabarés, F. Toral, J. Munilla, L. González, T. Puig, X. Obradors, *Perspectives in power applications of low and mainly high temperature superconductors: energy, transport and industry,* La Revista del Nuovo Cimento, vol. 48, pp. 435-536, 2025 (DOI: 10.1007/s40766-025-00071-6).
2. Z. Zhong and W. Wu, *High temperature superconducting closed-loop coils: constructures, key technologies, applications and future suggestions,* Supercond. Sci. Technol., vol. 38, 033002, 2025 (DOI: 10.1088/1361-6668/adb80b).

3. H. Zhang, Z. Wen, F. Grilli, K. Gyftakis, M. Mueller, *Alternating Current Loss of Superconductors Applied to Superconducting Electrical Machines,* Energies, vol. 14, p. 2234, 2021 (DOI:10.3390/en14082234).
4. S. Russenschuck, *Field computation for Accelerator Magnets: Analytical and Numerical Methods for Electromagnetic Design and Optimization*, Wiley-VCH Verlag GmbH & Co. KGaA, 2010 (DOI: 10.1002/9783527635467).
5. E. Pardo and F. Grilli, *Electromagnetic Modeling of Superconductors,* in: *Numerical Modeling of Superconducting Applications*, ed. B. Dutoit, F. Grilli, F. Sirois, World Scientific, 2023.
6. J.D. Jackson, *Classical Electrodynamics*, 3rd Edition, Wiley-VCH,1998, Ch.6, pp.237-294, ISBN 978-0-471-30932-1.
7. T. Tsang, *Classical Electrodynamics*, World Scientific,1997, Ch.4, pp.86-99, ISBN 981-02-3041-9, DOI: 10.1142/3388.
8. J.A. Stratton, *Electromagnetic Theory,* New York & London: McGraw-Hill, 1941, pp.226-229.
9. G. Bedrosian*, High-Performance Computing for Finite Element Methods in Low-Frequency Electromagnetics*, Progress in Electromagnetics Research, vol. 7, pp. 57-110, 1993.
10. P. Monk, *Finite Element Method for Maxwell's Equations,* Oxford, Clarendon, 2003.
11. J.L. Volakis, A. Chatterjee, L.C. Kempel, *Finite Element Method for Electromagnetics,* New York, Wiley-IEEE, 1998.
12. Q. Wang, *Practical design of magnetostatic structure using numerical simulations,* Singapore, John Wiley & Sons, 2013.
13. A.M. Campbell, *A direct method for obtaining the critical state in two and three dimensions,* Supercond. Sci. Technol., vol. 22, 034005, 2009 (DOI: 10.1088/0953-2048/22/3/034005).
14. G.P. Lousberg, M. Ausloos, C. Geuzaine, P. Dular, P. Vanderbemden, *Numerical simulation of the magnetization of high-temperature superconductors: a 3D finite element method using a single time-step iteration,* Supercond. Sci. Technol., vol. 22, 055005, 2009 (DOI: 10.1088/0953-2048/22/5/055005).
15. N. Nibbio, S. Stravrev, B. Dutoit, *Finite element method simulation of AC loss in HTS tapes with B-dependent E-J power law,* IEEE Trans. Appl. Supercond., vol. 11, pp. 2631-2634, 2001 (DOI: 10.1109/77.920408).
16. S. Stravrev, F. Grilli, B. Dutoit, N. Nibbio, E. Vinot, I. Klutsch, *Comparison of numerical methods for modeling of superconductors,* IEEE Trans. on Magnetics, vol. 38, pp. 849-852, 2002 (DOI: 10.1109/20.996219).
17. A. Stenvall, T. Tarhasaari, *An eddy current vector potential formulation for estimating hysteresis losses with FEM,* Supercond. Sci. Technol., vol.23, 125013, 2010 (DOI:10.1088/0953-2048/23/12125013).
18. V. Lahtinen, M. Lyly, A. Stenvall, T. Tarhasaari, *Comparison of three eddy current formulations for superconductor hysteresis loss modelling,* Supercond. Sci. Technol., vol. 25, 115001, 2012 (DOI: 10.1088/0953-2048/25/11/115001).
19. K. de Mare, M. Rang, R. Wesselink, *Predicting the critical current of a superconducting link,* in Proc. of the COMSOL Conference, Amsterdam, 2025.

20. A. Henning, M. Lindmayer, M. Kurrat, *Simulation Setup for Modeling the Thermal, Electric, and Magnetic Behavior of High Temperature Superconductors,* Phys. Procedia, vol. 36, pp. 1195-1205, 2012 (DOI: 10.1016/j.phpro.2012.06.201).
21. K. Berger, L. Quéval, A. Kameni, L. Alloui, B. Ramdane, F. Trillaud, L. M. Hell, G. Meunier, P. Masson, J. Lévêque, *Benchmark on the 3D Numerical Modeling of a Superconducting Bulk,* https://hal.science/hal-01548728v4, 2017.
22. M. Solovyov, F. Gömöry, *A-V formulation for numerical modelling superconductor magnetization in true 3d geometry*, Supercond. Sci. Technol., vol. 32, pp.115001-115010, 2019.
23. A. Musso, M. Breschi, P.L. Ribani, F. Grilli, *Analysis of AC Loss Contributions from Different Layers of HTS Tapes Using the A-V Formulation Model,* IEEE Trans. Appl. Supercond., vol. 31, pp.1-11, 2021 (DOI: 10.1109/TASC.2020.3033587).
24. M.D. Ainslie, H. Fujishiro, *Modelling of bulk superconductor magnetization*, Supercond. Sci. Technol., vol. 28, 053002, 2015 (DOI: 10.1088/0953-2048/28/5/053002).
25. K. Zhang, S. Hellmann, M. Calvi, T. Schmidt, L. Brouwer, *Magnetization Current Simulation of High Temperature Bulk Superconductors Using the ANSYS Iterative Algorithm Method*, IEEE Trans. Appl. Supercond., vol. 31, 9259084, pp.1-6, 2020 (DOI: 10.1109/TASC.2020.3037964).
26. G. dos Santos, B.M.O. Santos, F. Sass, F. G. dos Reis Martins, G.G. Sotelo, R. de Andrade Junior, *J-A formulation: A finite element methodology for simulating superconducting devices,* Superconductivity, vol. 6, 100049, 2023 (DOI: 10.1016/j.supcon.2023.100049).
27. S. Wang, H. Yong, Y. Zhou, *Numerical calculations of high temperature superconductors with the J-A formulation,* Supercond. Sci. Technol., vol. 36, 115020, 2023 (DOI: 10.1088/1361-6668/acfbbe).
28. Min Zhang, J-H. Kim, S. Pamidi, M. Chudy, W. Yuan, T.A. Coombs, *Study of second generation, high-temperature superconducting coil: Determination of critical current*, J. of Appl. Physics, 111, 083902, 2012 (DOI: 10.1063/1.3698317).
29. K.P. Thakur, A. Raj, E.H. Brandt, J. Kvitkovic, S. Pamidi, *Frequency-dependent critical current and transport ac loss of superconductor strip and Roebel cable,* Supercond. Sci. Technol., vol. 24, 065024, 11pp., 2011 (DOI: 10.1088/0953-2048/24/6/065024).
30. F. Gömöry and B. Klinčok, *Self-field critical current of a conductor with an elliptical cross-section*, Supercond. Sci. Technol., vol.19, pp.732-737, 2006.
31. F. Grilli, F. Sirois, V. M. R. Zermeno, M. Vojenčiak, *Self-consistent modeling of the $I_c$ of HTS Devices: How Accurate do Models Really Need to Be?* arXiv: 1405.5316v2 [cond-mat, supr-con] 2014.
32. M. Tomsic, M. Rindflesch, J. Yue, K. McFadden, J. Phillips, M.D. Sumption, M. Bhatia, S. Bohnenstiehl, E.W. Collings, *Overview of MgB2 Superconductor Applications,* Int. J. Appl. Ceram. Technol., vol.4, pp. 250-259, 2007.
33. A. Ballarino, *Development of superconducting links for the Hadron Collider machine,* Supercond. Sci. Technol. vol.27, 044024, 2014 (DOI:10.1088/0953-2048/27/4/044024).

34. F. Grilli, A. Chervyakov, V. Zermeno, A. Marian, G. Grasso, W. Goldacker, C. Rubbia, *Numerical modelling of $MgB_2$ conductors for high power AC transmission,* Physica C, vol. 504, pp.167-171, 2014 (DOI: 10.1016/j.physc.2014.04.037).
35. K. Vinod, R.G.A. Kumar, U. Syamaprasad, *Prospects for $MgB_2$ superconductors for magnet application,* Supercond. Sci. Technol., vol.20, R1, 2007 (DOI: 10.1088/0953-2048/20/1/R01).
36. T. Wang, F. Yang, D. Wang, Q. Guo, P. Zhang, *Advances in the performance and practicality of $MgB_2$ wires fabricated by internal magnesium diffusion,* Rev. Mat. Research, vol.2, 100325, 13 pages, 2026 (DOI: 10.1016/j.revmat.2026.1003250).
37. *COMSOL Multiphysics Reference Manual*, COMSOL AB, Stockholm, 2022.
38. F.X. Xiang, X.L. Wang, X, Xun, K.S.B. De Silva, Y.X. Wang, *Evidence for transformation from $\delta T_c$ to $\delta l$ pinning in MgB2 by graphene oxide doping with improved low and high field $J_c$ and pinning potential*, Appl. Phys. Lett., vol.102, 152601 (5pp), 2013.
39. H. Fujishiro, T. Naito and T. Yoshida, *Numerical simulation of the trapped field in $MgB_2$ bulk disks magnetized by field cooling*, Supercond. Sci. Technol., vol.27, 065019 (7pp), 2014.